\documentclass{article}
\usepackage{spconf,amsmath,graphicx}
\usepackage{adjustbox}
\usepackage{multicol,multirow}
\usepackage{enumitem}
\setlist{nosep, leftmargin=14pt}

\usepackage{mwe} % to get dummy images

\title{Ensemble Learning with Residual Transformer for Brain Tumor Segmentation}
\name{Lanhong Yao, Zheyuan Zhang, Ulas Bagci\thanks{This project is supported by the NIH funding: R01-CA246704 and R01-CA240639.}}
\address{Department of Radiology, Northwestern University, Chicago IL 60611, USA}
\begin{document}
%\ninept
%
\maketitle
\begin{abstract}
Brain tumor segmentation is an active research area due to the difficulty in delineating highly complex shaped and textured tumors as well as the failure of the commonly used U-Net architectures. The combination of different neural architectures is among the mainstream research recently, particularly the combination of U-Net with Transformers because of their innate attention mechanism and pixel-wise labeling. Different from previous efforts, this paper proposes a novel network architecture that integrates Transformers into a self-adaptive U-Net to draw out 3D volumetric contexts with reasonable computational costs. We further add a residual connection to prevent degradation in information flow and explore ensemble methods, as the evaluated models have edges on different cases and sub-regions. On the BraTS 2021 dataset (3D), our model achieves 87.6\% mean Dice score and outperforms the state-of-the-art methods, demonstrating the potential for combining multiple architectures to optimize brain tumor segmentation.

\end{abstract}
\begin{keywords}
Transformer, U-Net, Brain tumor segmentation, ensemble learning
\end{keywords}
\section{Introduction}
\label{sec:intro}
Brain tumor segmentation has been a challenging task in medical imaging research and plays a crucial role in diagnosis, prognosis, and determining treatment strategies. Magnetic resonance imaging (MRI) is the imaging modality of choice for brain tumor clinics due to its advantages~\cite{mri}: no radiation, better resolution of soft tissues compared with computed tomography (CT), and various imaging contrasts that provide rich diagnostic information via varying characteristics of the tissues highlighted by multi-modality MRI. For example, the spatial extent and volume of solid or active tumor regions, edema, and necrotic regions are important diagnostic and prognostic markers, and they can be measured by MRI. While tumor tissues may be readily detectable or visible at most times, accurate and reproducible auto-segmentation and characterization are still active research fields. This paper presents yet another deep learning-based segmentation method for brain tumor analysis using multi-modal MRIs with considerable advantages to the current state-of-the-art methods such as improving (1) the accuracy of segmentation results for complex shaped and textured tumors, (2) the robustness of the models and (3) the efficiency of segmentation systems (i.e., reducing the training time and computational burden).

The dataset used in this paper comes from Multi-modal Brain Tumor Segmentation Challenge (BraTS) 2021~\cite{baid2021rsna,menze2014multimodal,bakas2017advancing}, which includes four MRI modalities: T1-weighted (T1), T2-weighted (T2), fluid-attenuated inversion recovery (FLAIR), and contrast-enhanced T1-weighted (T1ce) MRI volumes. The segmentation task is defined as accurately generating contours of brain tumor sub-regions. % There are 3D images with multi classes, it is challenging  to apply conventional networks due to the mass computation brought by another dimension. Further, improving accuracy of all classes is an arduous task since the accuracy is already high given current quality of imaging date. An unbalanced distribution of classes is also likely to make it difficult to differentiate the sub-regions. 

\textbf{Previous works.} The U-Net~\cite{unet} architecture and its numerous modifications~\cite{nnunet,s3d,oktay2018attention,luu2022extending} have achieved the state-of-the-art results in medical image segmentation. Skip connections between the encoder and decoder enable the network to capture details from low-level layers for accurate localization of object of interest. However, due to limited kernel size, long-range dependency is often lost through this process. This is rather important in biomedical segmentation tasks because it is difficult to capture more complex relations in some tumors with complex shapes and texture. 

To address this issue, many efforts of introducing the self-attention mechanism have recently been made~\cite{valanarasu2021medical}. For instance, TransUNet~\cite{TransUnet} takes image patches from U-Net and provides evidence that Transformers have the potential in making encoders stronger. U-Net Transformer~\cite{UnetTrans} and SwinUNet~\cite{SwinUnet} both apply the transformer blocks during the low-level encoding-decoding process to model long-range dependencies. These models have shown the ability to extract global context; therefore, leading to a better performance on medical image segmentation. Yet, the transformer brings a heavy computational and memory burden.

Despite the benefits of Transformers, the complexity required by its mechanism can largely impede the feasibility of its application on 3D datasets. The aforementioned approaches are applied mainly on 2D medical images~\cite{mm}, which leaves room for the further explorations of Transformers. However, for 3D images, since the transformer requires quadratic complexity calculations, the one more dimension of the dataset would increase the computation so dramatically that it cannot be naively integrated into the network architecture. One potential solution is to introduce a \emph{linear transformer} for segmentation, however, previous research also shows there is a  performance drop when making the transformer linear \cite{zhang2022dynamic}.

To this end, we propose a novel architecture by integrating Transformers into the nnU-Net architecture solely in the bottleneck layer of feature extraction so as to alleviate the heavy computation issue on 3D datasets. Our models are expected to combine the benefits of the self-attention mechanism and nnU-Net, which are to learn long-range dependencies, to reserve the details from low-level layers in the decoding process, and to auto-generate network configurations. To avoid hierarchical complexity brought by additional layers~\cite{residual}, we also add residual connection on the Transformer block to further stabilize the model performance. 

Our contributions can be summarized as:
\begin{itemize}
% \item We linearize Transformers to speed up its computation approximately 100 times better than conventional Vision Transformer (ViT).
\item We design an effective architecture combining the Transformer algorithm with a self-adaptive U-Net and include residual connections on the Transformer to avoid possible degradation in information flow. 
%To our knowledge, there has been no attempts to examine such combined network architectures in the literature. 
\item Further, we manage to apply our algorithm in 3D volumetric space instead of the conventional slice-by-slice approach.
\item We demonstrate a great potential in ensemble learning of the proposed architectures for further enhancement in the segmentation performance. 
\item The proposed design shows improved robustness against poor-quality images, and the efficacy for brain tumor segmentation on 3D datasets and outperforms the state-of-the-art approaches in this field.
\end{itemize}

\section{Methodology}
\label{sec:format}
This study is based on nnU-Net, an extension of U-Net that automatically generates network configurations based on the dataset characteristics~\cite{nnunet}. In our implementation, the parameters, including target spacing, patch size, batch size, downsampling strides, and convolutional kernel sizes, are self-adaptive and adjusted based on the specific dataset. To capture long-range contextual interactions, we introduce a self-attention mechanism by integrating a transformer block into the deepest layer of the encoder, where the dimensions of the input sequences are reduced to a size that can be easily handled by the transformer. We also include a residual connection to prevent information degradation. Figure 1 shows the architecture of our proposed model, illustrating the flow of information through the network, and highlighting the locations of the transformer block and residual connection.

\begin{figure}[htb]
    \begin{minipage}[b]{1.0\linewidth}
      \centering
      \centerline{\includegraphics[width=9cm]{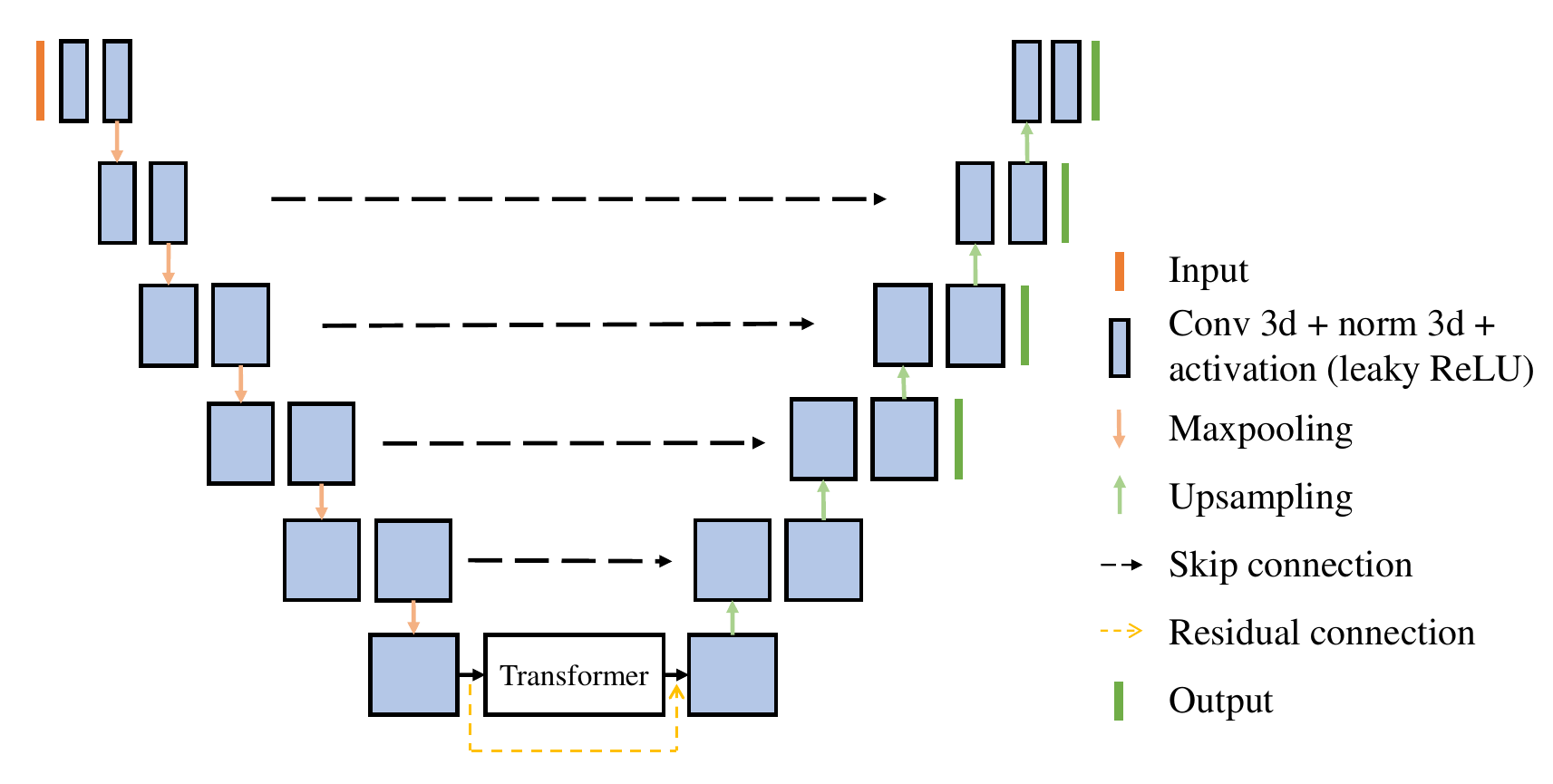}}
    \end{minipage}
\caption{The network architecture of the segmentation engine with the transformer block nested at the bottleneck layer.}
\label{fig1:network}
\end{figure}

\subsection{U-Net architecture with transformer}
% \subsubsection{Down-sampling}
\quad ~\textbf{Down-sampling.}
The input to our model is a 3D image with dimensions of 240 x 240 x 155 (height x width x depth), composed of four distinct channels derived from various MRI contrasts. Following pre-processing (i.e., inhomogeneity correction, denoising, and intensity standardization), the images are conveyed to the encoder for down-sampling. Each layer of the encoder comprises two consecutive blocks of stacked convolutional layers, incorporating 3D convolution, dropout, 3D instance normalization, and leaky ReLU activation in a sequential manner. The encoder extracts features from image patches that are subsequently utilized as input sequences for the Transformer block.

% \subsubsection{Transformer block}
\textbf{Transformer block.}
The Transformer block acts as the final component of the encoder, performing a sequence-to-sequence operation. It takes in non-overlapping encoded patches and transforms them into vectors. It repeats the self-attention layer $n=8$ times for multi-head attention. The self-attention~\cite{attention} mechanism can be written as:
\begin{equation}
    Attention (\textbf{Q}, \textbf{K}, \textbf{V}) = softmax (\frac{\textbf{QK}^T}{\sqrt{d}})\textbf{V},
\end{equation}

\noindent where the manipulation of queries (Q), keys (K), and values (V) requires O(N²) computational complexity. After being encoded down to the bottleneck layer, the sequence dimension could be reduced dramatically. Thus, on a 3D dataset, the transformer block placed in the last layer can alleviate the computational burden and can make it feasible to include self-attention in the overall network. The Transformer later outputs encoded image patches into the decoder, which reshapes them back to the original input size.

\begin{figure}[ht]
    \begin{minipage}[b]{1.0\linewidth}
      \centering
      \centerline{\includegraphics[width=8cm]{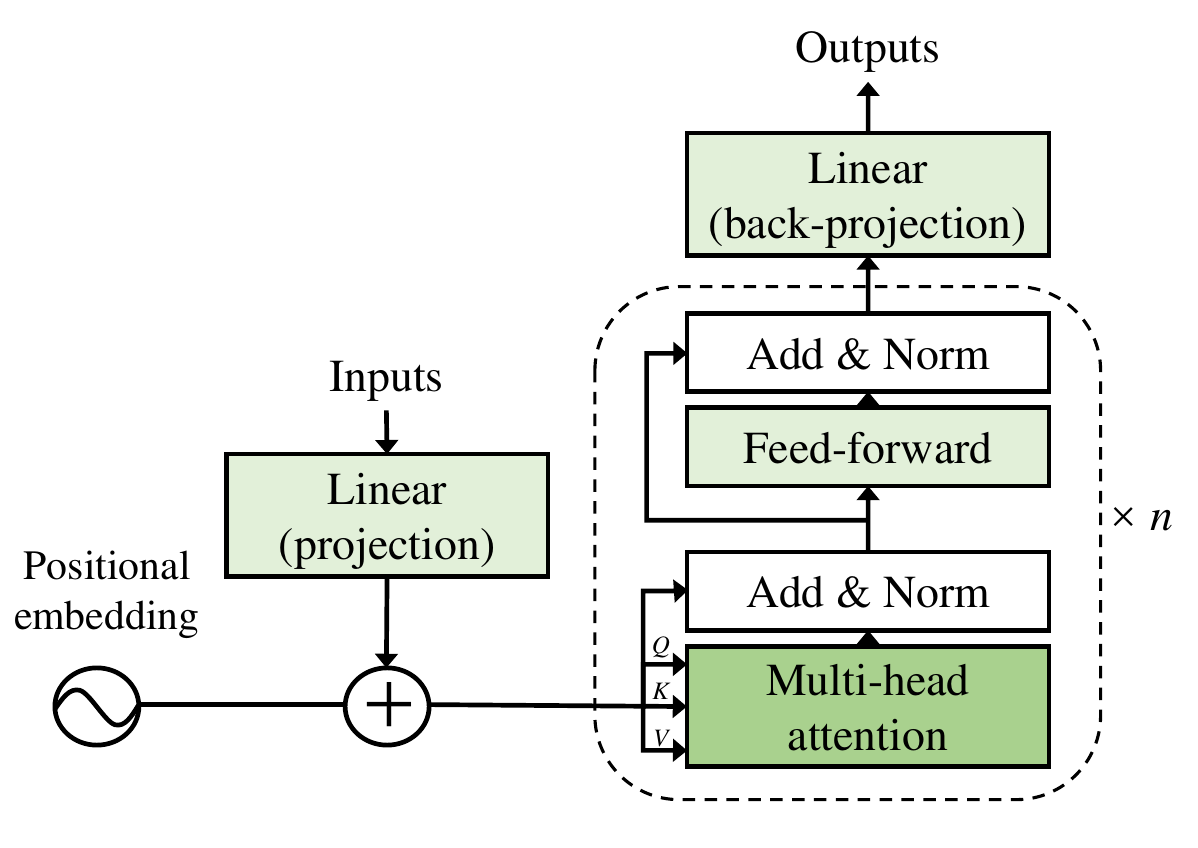}}
    \end{minipage}
\caption{Zoom-in architecture of the transformer block.}
\label{fig:transformer}
\end{figure}

% \subsubsection{Residual connection}
\textbf{Residual connection.}
To avoid a fracture in information from previous layers and the difficulty in training complex models, we add a residual connection to the Transformer block. This allows the information to be propagated directly to the designated last layer. By doing so, we seek to ensure the integrity of the information flow while mitigating potential issues such as gradient explosion. We compare this model to one without a residual connection.

% \subsubsection{Up-sampling}
\textbf{Up-sampling.}
During the up-sampling process, the decoder utilizes transposed convolutions to expand the image size and move up to low-level layers until the size matches the original input image. At each layer, the decoder concatenates input features with image features from the corresponding encoder level through a skip connection. It preserves crucial details in the final segmentation map. To further prevent information loss in the decoding process, we incorporate deep supervision. Overall, this up-sampling strategy enhances the accuracy and robustness of the proposed architecture.

% \subsection{Implementation details}
\subsection{Training and testing}
We evaluate all our models on the BraTS2021 dataset. This dataset contains fully-annotated, multi-institutional, multi-parametric brain tumor MRI, covering diverse degrees of gliomas. There are 1,470 samples in the dataset, each of which has four MRI modalities: T1, T2, FLAIR, and T1ce. We split them into the training and validation set of 1,251 samples, and the test set of 219 samples. We use 5-fold cross-validation and optimization to train our models, where the loss function is defined as the sum of Dice loss and cross-entropy loss. Moreover, to assess the performance of our models, we conduct a blind-testing on the unseen validation dataset through an online evaluation provided by the BraTS challenge organizers.

\begin{figure}[htb]
    \begin{minipage}[b]{1.0\linewidth}
      \centering
      \centerline{\includegraphics[width=9.0cm]{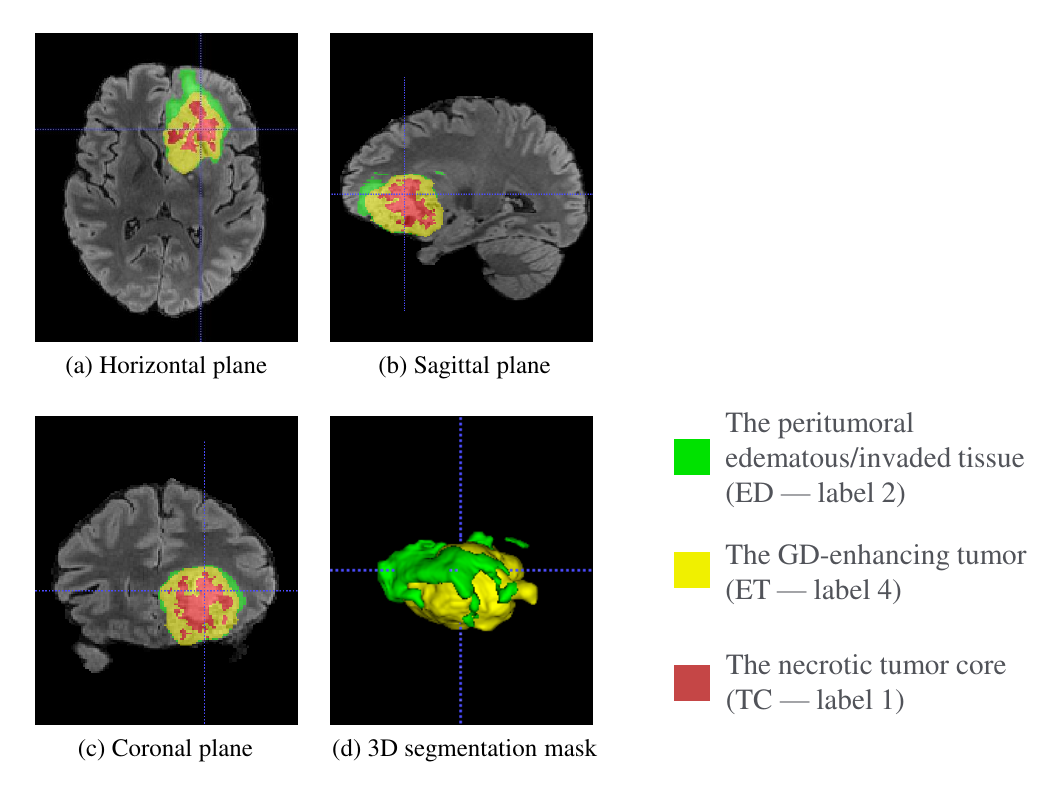}}
    \end{minipage}
\caption{Typical labels in expert segmentation on a sample from the training set, projected on FLAIR.}
\vspace{-4mm}
\label{fig:labels}
\end{figure}

\begin{figure}[htb]
    \begin{minipage}[b]{1.0\linewidth}
      \centering
      \centerline{\includegraphics[width=8cm]{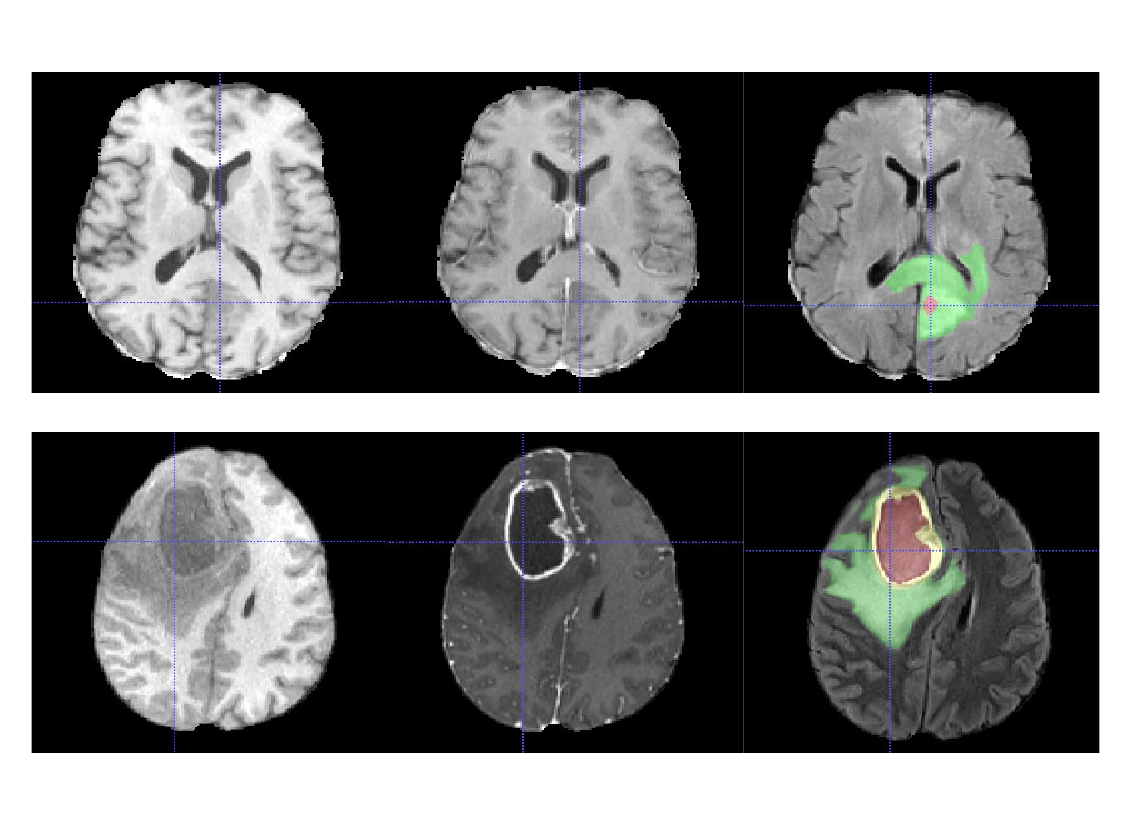}}
    \end{minipage}
\caption{Segmentation of tumors is illustrated (each row indicates one example). Representatives of failure (left) and acceptable (right) segmentation predictions for a given image (middle). Left to right, modalities are T1, T1ce, and FLAIR with a half-transparent segmentation map.}
\label{fig:examples}
\vspace{-8mm}
\end{figure}

Figure~\ref{fig:labels} illustrates an example annotation of a single subject's FLAIR image with four segmentation labels: peritumoral edema (ED), GD-enhancing tumor (ET), necrotic tumor core (TC), and background. A whole tumor (WT) refers to the complete extent of the disease, entailing TC and ED. When comparing T1ce to T1 images, ET shows hyper-intensity while TC shows hypo-intensity, as can be observed in Figure \ref{fig:examples}. Generally, all four labels are present in the expert manual annotation following a certain topological relationship, where ED is on the outside and TC is the innermost, but it varies from patient to patient. Some samples only have one or two labels (when there is no ET or TC to be found) due to different tumor development stages, which may affect the segmentation performance of certain models. This will be further discussed later.

\section{Results}
\label{sec:pagestyle} 
\subsection{Single model performance}
We evaluate the performance of four different models on the BraTS2021 dataset. The models include nnU-Net as the baseline, E1D3~\cite{e1d3} as a comparison to the state-of-the-art, and our proposed models transformer-nested nnU-Net (UNet+T) and its variation with residual connection (UNet+T+R). E1D3 is a variation of the U-Net that uses one encoder and three decoders for separate sub-regions of the brain tumor. This architecture may give E1D3 some advantages on certain samples that require examining sub-regions independently, therefore making it a good candidate for the later ensemble.

The evaluation metrics used in this study were the Dice coefficient score and Hausdorff Distance (HD). Table \ref{tab1} provides the evaluation results on sub-region performance for all four models. The results show that UNet+T has a slightly better Dice score on ET and TC, and a better HD score on WT. UNet+T+R demonstrates significant improvement in the prediction of ET and better HD scores on ET and TC. As a result, our models demonstrate the best overall performance as evaluated by both mean Dice and mean HD.

Furthermore, nnU-Net is observed to be robust to poor image quality and displacement noise, as it gives the highest Dice score for WT. We also observe that E1D3 performs better on the ET region for many selected samples that contains less than or equal to two labels in the segmentation map, excluding the background. This observation inspires us to \emph{ensemble} the results generated by different models to combine their advantages on sub-region segmentation. 

\subsection{Ensemble models}
Table \ref{tab2} presents the results of four different ensemble strategies. The first three methods involves performing pixel-wise operations on the segmentation maps from the four models, using mode, mean, and median, respectively. We should note that the weight of each model is not equal, as we double the weight of the UNet+T model based on its better overall performance and empirical evidence. For the threshold ensemble, we set a threshold of TC volume smaller than 60 and ET volume larger than 60 to automatically select subjects to be predicted by the E1D3 model, and the remaining subjects are predicted using the UNet+T model. This method is based on the observation that E1D3 performs better for subjects meeting these criteria, which may suggest fewer than four labels in the ground truth. Therefore, this method is only an ensemble of two models: E1D3 and UNet+T. The threshold method shows the best overall performance, both in terms of Dice score and HD, suggesting that this criterion helps improve our models.

\begin{table}
    \caption{Prediction results for 4 models on the BraTS 2021 validation dataset (unseen). Bold texts represent the best scores for each metric. HD refers to Hausdorff Distance. T and R stand for transformer block and residual connection around it, respectively.}
    \label{tab1}
    \begin{center}
    \begin{adjustbox}{width=\columnwidth}
        \small
        \begin{tabular}{|c|c|c|c|c|c|c|c|c|}
            \hline
            \multirow{2}{*}{\textbf{Methods}} & 
            \multicolumn{4}{ c |}{\textbf{Dice (\%)}}  & 
            \multicolumn{4}{ c |}{\textbf{HD (mm)}}  \\ 
            \cline{2-9}
            & \textbf{ET} & \textbf{TC} & \textbf{WT} & \textbf{Mean}
            & \textbf{ET} & \textbf{TC} & \textbf{WT} & \textbf{Mean}\\
            \hline
            nnU-Net        & 80.5 & 87.5 & \textbf{92.7} & 86.9 & 24.66 & 7.62 & 3.60 &11.96\\ \hline
            E1D3           & 81.6 & 80.6 & 91.5 & 84.6 & 18.46 & 17.65& 4.73 &13.61\\ \hline
            UNet+T     & 82.0 & \textbf{87.8} & 92.6 & 87.5 & 18.13 & 9.27 & \textbf{3.49} & 10.30\\ \hline
            UNet+T+R & \textbf{82.2} & 87.7 & 92.5 & \textbf{87.5} & \textbf{16.24} & \textbf{7.53} & 3.73 &\textbf{9.17}\\ \hline
        
        \end{tabular}
        \end{adjustbox}
    \end{center}
\end{table}

\begin{table}
\caption{Prediction results for ensemble models on the BraTS 2021 validation dataset (unseen). Bold texts represent the best scores for each metric.
}
\label{tab2}
\begin{center}
\begin{adjustbox}{width=\columnwidth}
\small
\begin{tabular}{|c|c|c|c|c|c|c|c|c|}
% \begin{tabular}{|c|*{8}{p{1.3cm}|}}
    \hline
    \multirow{2}{*}{\textbf{Methods}}& \multicolumn{4}{ c |}{\textbf{  Dice (\%)  }}  & \multicolumn{4}{ c |}{\textbf{HD (mm)}}  \\ 
    \cline{2-9}
    & \textbf{ET} & \textbf{TC} & \textbf{WT} & \textbf{Mean}
    & \textbf{ET} & \textbf{TC} & \textbf{WT} & \textbf{Mean}\\
    \hline
    Mode      & 81.6 & 87.6  & 92.6 & 87.3 & 19.72 & 9.26 &3.61 &10.86\\ \hline
    Average       & \textbf{82.4} & 80.6  & 92.2 & 85.1 & \textbf{12.87} & 16.63 &5.27 &11.59\\ \hline
    Median    & 82.0 & 87.7  & \textbf{92.6}  & 87.4 & 18.00 & 9.20 &3.61 &10.27\\ \hline
    Threshold & 82.0 & \textbf{88.3}  & 92.6  & \textbf{87.6} & 18.13 & \textbf{7.56}&\textbf{3.57} &\textbf{9.75}\\ \hline
    % "Manual"  & 84.0 & 88.3  & 92.8  & 88.4 & 11.22 & 5.83&3.41\\ 
    % \hline
    \end{tabular}
\end{adjustbox}
\end{center}
\end{table}

We also manually pick the best model to apply for each case, but the results are not compared in the table because it is not an automatic process. However, it has achieved the highest Dice (ET 84.0\%, TC 88.3\%, WT 92.8\%) and lowest HD (ET 11.17mm, TC 7.51mm, WT 3.40mm) across all sub-regions, as well as the mean Dice (88.4\%) and mean HD (6.82mm). It indicates that the four architectures evaluated in this paper have the potential to improve combined accuracy significantly if a more complex ensemble method were specified. All these ensemble methods have improved accuracy on selected sub-regions, and the threshold method gives the best overall Dice score and HD among automatic methods.

\section{Discussion}
\label{sec:typestyle}

We present a residual Transformer integrated nnU-Net architecture for brain tumor segmentation. Simple yet effective, both models with Transformer nested demonstrated impressive performance on 3D multi-modality datasets and have surpassed the state-of-the-art algorithms, including nnU-Net itself. We also explore ensemble models in an effort to combine their advantages on sub-regions and different types of subjects. Although the ensemble methods are not complex and the threshold is empirical, we have still observed notable improvements in segmentation performance.

We observe that the performance of the model is affected by various aspects of the subject, such as the number of sub-regions, the percentage of each class, the clarity of the boundary, and other unidentified factors. To test this hypothesis, we manually pick the best-performing model for each subject, and it achieves the best Dice scores among all models across metrics. This observation suggests that human-in-the-loop decision-making, as is common in current clinical practice, can further push the boundary of current segmentation performance.

In the future, we plan to investigate better ensemble methods and select appropriate models for combination. The potential demonstrated in our results is convincing and has the potential to be applied to other medical imaging tasks. Overall, our work offers a promising ensemble approach to brain tumor segmentation, opening up new possibilities for the integration of Transformers in medical image analysis.

\bibliographystyle{IEEEbib}
\bibliography{strings}

\end{document}